\documentclass[amsmath,amssymb,11pt,aps,a4paper]{revtex4-2}
\usepackage{amsfonts}
\usepackage{amsmath}
\usepackage[utf8]{inputenc} 
\usepackage{graphicx}
\usepackage{float}
\usepackage{natbib}
\usepackage{xcolor}
\usepackage{dcolumn}
\usepackage[english]{babel}
\usepackage{soul}
\usepackage{mathtools}

\newcommand{\be}{\begin{equation}}
\newcommand{\ee}{\end{equation}}

\newcommand{\dth}[1]{\dot{\theta}_{#1}}

\begin{document}
	
\title{Exact solutions of the Kuramoto model with asymmetric higher order interactions of arbitrary order}

\author{Guilherme S. Costa$^1$}
\author{Marcel Novaes$^{2,3}$}	
\author{Marcus A.~M.~de Aguiar$^{1,3}$}

\affiliation{$^1$ ICTP South American Institute for Fundamental Research \& Instituto de Física Teórica - UNESP, 01140-070, São Paulo, Brazil }
\affiliation{$^2$Instituto de F\'isica, Universidade Federal de Uberl\^andia, 38408-100, Uberl\^andia, MG, Brazil}
\affiliation{$^3$Instituto de F\'isica Gleb Wataghin, Universidade Estadual de Campinas, 13083-970, Campinas, SP, Brazil}

\begin{abstract}

Higher order interactions can lead to new equilibrium states and bifurcations in systems of coupled oscillators described by the Kuramoto model. However, even in the simplest case of 3-body interactions there are more than one possible functional forms, depending on how exactly the bodies are coupled. Which of these forms is better suited to describe the dynamics of the oscillators depends on the specific system under consideration. Here we show that, for a particular class of interactions, reduced equations for the Kuramoto order parameter can be derived for arbitrarily many bodies. Moreover, the contribution of a given term to the reduced equation does not depend on its order, but on a certain effective order, that we define. We give explicit examples where bi and tri-stability is found and discuss a few exotic cases where synchronization happens via a third order phase transition.
		
\end{abstract}

\maketitle
	
\section{Introduction}
\label{intro}

In systems composed of multiple particles, interactions may go beyond pairwise relations and involve the collective action of groups of agents that cannot be decomposed. A classic example is collaboration networks, where more than two people can participate in a project or coauthor a paper \cite{vasilyeva2021multilayer}. In physics, the Einstein–Infeld–Hoffmann equations of motion, which incorporate small general-relativistic effects into many-body newtonian mechanics, lead to gravitational forces that are proportional to the product of several different masses \cite{landau,einstein}. Other examples can be found in neuroscience \cite{ganmor2011sparse,petri2014homological,giusti2015clique,reimann2017cliques,sizemore2018cliques,majhi2025144}, 
ecology \cite{grilli2017higher,ghosh2024chimeric}, biology \cite{sanchez2019high} and social sciences \cite{benson2016higher,de2020social,alvarez2021evolutionary}. Higher order interactions have particularly important consequences in the propagation of epidemics \cite{iacopini2019simplicial,jhun2019simplicial,vega2004fitness} and synchronization of coupled oscillators \cite{berec2016chimera,skardal2019abrupt,skardal2020higher,dai2021d,sarika2024,biswas2024symmetry,sayeed2024global,muolo2024phase,tanaka2011multistable,bick2016chaos,battiston2020networks,dutta2023impact,leon2024higher}.  

Pairwise interactions can be described by a network, where nodes represent agents, and links indicate if the corresponding pair of agents interact. For higher order interactions, networks are replaced by simplicial complexes or hypergraphs which, besides nodes and links, also contain triangles, tetrahedra, and higher order cliques. However, higher-order interactions can take different functional forms, depending on how each node on the simplex contributes to the dynamics. Importantly, these different forms can have profound effects on the global properties of the system. 

In this work we consider a particular class of higher-order interactions on the Kuramoto model that can be treated analytically to all orders. We show that the qualitative effect of each order does not depend on the order itself, but on an effective order $p$ which organizes the reduced equation for the order parameter and controls the phase transitions. 

The original Kuramoto model describes a set of $N$ oscillators that interact with each other according to the equation
\be \dot{\theta}_i = \omega_i + \frac{K}{N} \sum_{j=1}^N \sin{(\theta_j-\theta_i)},\ee 
where $K$ is a coupling constant and $\omega_i$ is the natural frequency of the $i$-th oscillator. If the natural frequencies are drawn from a smooth unimodal distribution with zero average, $g(\omega)$, a continuous (second order) phase transition to synchronization takes place at $K=K_c=2/(\pi g(0))$ in the large-$N$ limit. For $K \gtrapprox K_c$, a small group of oscillators is synchronized and, as $K$ increases beyond $K_c$, the group of synchronized oscillators increases, encompassing all particles as $K \rightarrow \infty$. 

According to phase reduction methods \cite{ashwin2016hopf,leon2019phase}, three-body interactions proportional to $\sum_{j,k} \sin{(2\theta_j-\theta_k -\theta_i)}$ or $\sum_{j,k} \sin{(\theta_j+\theta_k -2\theta_i)}$ can be added to the equation of motion of $\dot{\theta}_i$ \cite{battiston2020networks}.  Both interactions go to zero when the corresponding oscillators on the 2-simplex (triangle) have the same phase. The first form is asymmetric with respect to the exchange of the summed indices $j$ and $k$, while the second is symmetric. Similarly, symmetric four-body interactions are proportional to $\sum_{j,k,l} \sin{(\theta_j+\theta_k+\theta_l -3\theta_i)}$, while asymmetric interactions can have, for example, the form $\sum_{j,k,l} \sin{(\theta_j-\theta_k+\theta_l -\theta_i)}$ (all examples in this paragraph preserve rotational symmetry). 

These forms of interaction have very different implications for the dynamics of the Kuramoto model. Symmetric terms have been studied in \cite{tanaka2011multistable,skardal2019abrupt,dai2021d,leon2024higher} and result in multi-stability. Dai et al \cite{dai2021d} have also considered the effects of symmetric interactions of arbitrary order in the multidimensional Kuramoto model. Asymmetric interactions, on the other hand, were considered in \cite{skardal2020higher,dutta2023impact,fariello2024third,suman2024finite} and were shown to give rise to bi-stability. 

Other types of interaction have also been discussed in the literature, such as $\sum_j \sin(\theta_j+\theta_i)$, which breaks rotational symmetry \cite{manoranjani2023diverse,buzanello2022matrix,de2023generalized}, or $\sum_j \sin(2(\theta_j-\theta_i))$, representing higher harmonics of pair interaction \cite{holzel2015stability,nishikawa2004capacity,nishikawa2004oscillatory}. 
Taking multiple forms of interactions into account simultaneously is generally difficult due to the complexity and non-linearity of the equations. Systems with only one such term, such as pure symmetric three-body interactions, were considered in \cite{skardal2019abrupt} and \cite{dai2021d}, but numerical simulations must be employed when 2- and 3-body interactions are taken into account simultaneously. 

Among the many types of asymmetric interactions, those of the form $ \sin (\vec{\theta}\cdot \vec{c} - \theta_i)$, where $\vec{c}$ is a vector of integer entries, with $c_i \neq 0$ and satisfying $\sum_{j} c_j = 1$, play a special role, as they can be treated with the Ott-Antonsen ansatz and provide exact solutions for the dynamics of the system. Skardal and Arenas \cite{skardal2020higher} have considered 3 and 4-body asymmetric interactions of this type, with coupling constants $K_2$ and $K_3$, respectively, acting along with pairwise interactions with coupling intensity $K_1$. They showed that if $K_2$ or $K_3$ are large enough, they can lead to bi-stability, promoted by a saddle-node bifurcation, and the appearance of a first-order phase transition \cite{skardal2020higher}. Moreover, the dynamics of the system depends on $K_2$ and $K_3$ only through their sum, $K_{23}=K_2+K_3$. This result hints that the effective contribution of these terms is not related to their order, but to other properties of the interaction. Here we generalize this result for the above class of asymmetric interactions and identify the effective order that controls the role of the interaction in the dynamics. We use the OA ansatz to derive exact equations for the order parameter of the system and construct bifurcation diagrams of the model for some combinations of higher order terms.

\section{Derivation of reduced equations}
\label{ahoi}

\subsection{Specifying the higher order interactions}

We start by considering a set of $N$ oscillators interacting according to the Kuramoto model where interactions occur in pairs and also contain higher order (many body) terms. We write, rather informally at this point,
\be \dot{\theta}_i = \omega_i + \frac{K}{N} \sum_{j=1}^N \sin{(\theta_j-\theta_i)} + higher \; order \; terms.
\label{inf} \ee 
We now make the key assumption that the $(\beta+1)$-body contribution (corresponding to a $\beta$-simplex) to the evolution equation  has the form
\begin{eqnarray}
	\dfrac{K_{\vec{c}}}{N^\beta} \sum_{\{j_1,j_2,...,j_\beta = 1\}}^N \sin (c_1 \theta_{j_1} + c_2 \theta_{j_2} + \dots 
      + c_\beta \theta_{j_\beta} - \theta_i ).
\end{eqnarray}
The $N^\beta$ in the denominator is important to make everything well defined in the limit $N\to \infty$.

The peculiarity of this form of interaction is that the focal particle $i$ contributes with fixed coefficient $-1$, which is crucial to the application of the OA ansatz to the full equation. The other coefficients are integers and must satisfy the relation
\begin{equation}
	\sum_{i=1}^\beta c_i = 1
\end{equation}
to preserve rotational symmetry and to ensure that the interaction vanishes when the oscillators in the simplex are synchronized. In compact form we can rewrite the term as
\begin{eqnarray}
	\dfrac{K_{\vec{c}}}{N^\beta} \sum_{\vec{j}}^N \sin (\vec{\theta}\cdot \vec{c} - \theta_i)
	\label{kuramoto}
\end{eqnarray}
where $\vec{c}$ is a vector of size $\beta$ with integer entries and $K_{\vec{c}}$ is the coupling constant for the specific choice of $\vec{c}$ and order $\beta$. The {\it effective order} of the interaction is now defined as the sum of the positive coefficients of $\vec{c}$:
\begin{equation}
    p = \sum_{i=1}^\beta c_i \, \Theta(c_i)
    \label{eqp}
\end{equation}
where $\Theta(x)$ is the Heaviside function. It follows that the $\ell_1$-norm of $\vec{c}$ is
\begin{equation}
	\sum_{i=1}^\beta |c_i| = 2p-1.
    \label{eqpp}
\end{equation} 

Therefore, the general equation for the Kuramoto model with higher-order interactions of this specific functional form can be written as a sum over several terms like Eq.(\ref{kuramoto}), each one representing a $(\beta+1)$-body term with their own coupling constant $K_{\vec{c}}$, oscillator vector  $\vec{\theta}$ and coefficient vector $\vec{c}$.

\subsection{Ott-Antonsen ansatz}

Defining the generic Daido order parameter, for $m> 0$, as
\begin{equation}
	z_m = r_m e^{i \psi_m} \equiv \dfrac{1}{N} \sum_{j=1}^{N} e^{i m \theta_j},
    \label{eqdaido}
\end{equation}
we can write the arbitrary higher-order asymmetric term, Eq \eqref{kuramoto}, as
\begin{align*}
	\dfrac{1}{N^\beta}  \sum_{\vec{j}} \sin (\vec{\theta}\cdot \vec{c} - \theta_i) &= \text{Im}\left[\dfrac{\sum_{\vec{j}} e^{i(\vec{\theta} \cdot \vec{c}-\theta_i)}}{N^\beta}\right] = \text{Im}\left[\sum_{\vec{j}}\dfrac{ \prod_l e^{ic_l\theta_{j_l}}}{N^\beta} e^{-i\theta_i}\right] \\
   &= \left( \prod_l r_{|c_l|} \right)    \sin \left(\sum_l \psi_{|c_l|} \dfrac{c_l}{|c_l|} - \theta_i \right)
\end{align*}  
To simplify the notation and using the fact that the Ott-Antonsen ansatz can be applied separately to each term, let us consider a $(\beta+1)$-body arbitrary interaction written as 
\begin{equation}
	\dth{i} = \omega_i + k \left( \prod_l r_{|c_l|} \right)     \sin \left(\sum_l \psi_{|c_l|} \dfrac{c_l}{|c_l|} - \theta_i \right).
\end{equation}

In the limit $ N \to \infty$, the oscillators may be described by a continuous density $f(\omega,\theta,t)$ that determines the probability of finding oscillators of natural frequency $\omega$ at position $\theta$ at time $t$. Ott and Antonsen \cite{Ott2008} realized that, if the density is given by
\begin{equation}
    f(\omega,\theta,t)=g(\omega)\frac{1}{2\pi} \frac{1-|\alpha|^2}{|1-\bar{\alpha}e^{i\theta}|^2},
\end{equation}
with $g(\omega)$ multiplied by the Poisson kernel, the Daido parameters satisfy
\begin{equation}
    z_m(t)=\int [\bar{\alpha}(\omega,t)]^mg(\omega)d\omega.
\end{equation}

In particular, if $g(\omega)$ is a Lorentzian distribution of mean $\omega_0$ and half-width $\Delta$, the above integral can be computed by the methods of residues and results in 
\begin{equation}
    z_m = \bar{\alpha}^m(\omega_0-i\Delta).
\end{equation}
Moreover, the density satisfies the continuity equation and maintains this functional form at all times, provided $\alpha(t)$ evolves according to \cite{skardal2020higher}
\begin{equation}
	\label{eq::dyneta}
	\dot{\alpha} =  -i\omega \alpha  + \dfrac{H}{2} - \dfrac{H^*}{2}\alpha^2
\end{equation}
where
\begin{equation}
	H = k \prod_{l=1}^{\beta} r_{|c_l|}\text{exp}\left({i\psi_{|c_l|} \frac{c_l}{|c_l|}}\right) = k|z|^{2p-2}z.
\end{equation}
Here we used Eqs.(\ref{eqp}) and (\ref{eqdaido}) for the effective order of the interaction and Daido order parameters. As we shall see in the following, $p$ organizes the reduced equation for the order parameter of the Kuramoto model and plays an important role in the bifurcation diagrams.

%
%
Calculating (\ref{eq::dyneta}) at $\omega=\omega_0 - i \Delta$ we obtain
\begin{equation}
    \label{eq::dyneta2}
	\dot{z}=  i\omega_0 z  - \Delta z + k\dfrac{|z|^{2p-2}z}{2} - k\dfrac{|z|^{2p}z}{2},
\end{equation}
or, in polar coordinates,
\begin{equation}
	\dot{r} = -\Delta r + \dfrac{(1-r^2)kr^{2p-1}}{2}
    \label{eqrfinal}
\end{equation}
\begin{equation}
	\dot{\psi} = \omega_0.
    \label{eqpsifinal}
\end{equation}

By changing to a reference system that rotates along with $\psi$, we can set $\omega_0 = 0$ without loss of generality, therefore eliminating Eq.(\ref{eqpsifinal}). For an arbitrary number of higher order terms, the order parameter equation can be written as
\begin{equation}
	\dot{r} = -\Delta r + \dfrac{1}{2}(1-r^2)\left(\sum_{p}k_p r^{2p-1} \right).
    \label{rdot}
\end{equation}
This equation can be simplified slightly by introducing $\rho=r^2$. Then 
\begin{equation}
	\dot{\rho} = -2\Delta \rho +(1-\rho)\left(\sum_{p}k_p \rho^{p} \right).
    \label{rho}
\end{equation}

In these equations, $k_p$ is the additive contribution of all terms in the original equation for which the $\ell_1$-norm of $\vec{c}$ is $2p-1$, or, equivalently, for which the sum of positive entries in $\vec{c}$ is $p$. For example, if
\begin{eqnarray}
	\dot{\theta}_i &=& \omega_i + \frac{K_1}{N} \sum_{j=1}^N \sin{(\theta_j-\theta_i)} + 
	\frac{K_2}{N^2} \sum_{j,k=1}^N \sin{(2\theta_j-\theta_k -\theta_i)} \nonumber \\
	&+& \frac{K_3}{N^3} \sum_{j,k,m=1}^N \sin{(\theta_j-\theta_k +\theta_m-\theta_i)} + 
    \frac{K_4}{N^4} \sum_{j,k,m,l=1}^N \sin{(2\theta_j-\theta_k +\theta_m - \theta_l -\theta_i)} \nonumber \\
    &+& \frac{K_5}{N^5} \sum_{j,k,m,l,n=1}^N \sin{(\theta_j+\theta_k +\theta_m - \theta_l - \theta_n -\theta_i)}+ \frac{K_6}{N^2} \sum_{j,k=1}^N \sin{(3\theta_j-2\theta_k -\theta_i)}
\end{eqnarray}
then $k_1=K_1$, $k_2 = K_2 + K_3$, $k_3=K_4 + K_5+K_6$ and the equation for the order parameter becomes
\begin{equation}
	\dot{r} = -\Delta r + \dfrac{1}{2}(1-r^2) \left(k_1 r + k_2 r^3 + k_3 r^5 \right).
    \label{eqr}
\end{equation}

\section{Equilibrium points and bifurcation diagrams}
\label{equilibrium}

\subsection{General properties}

The point $r^* = \rho^*=0$ is always an equilibrium solution of Eqs. (\ref{rdot}) and (\ref{rho}), being responsible for the pitchfork bifurcation, which occurs at $k_1 = 2\Delta$ and is independent of other $k_n$. The exact form of the bifurcation depends on the sign of the third order derivative in $r$, being supercritical if it is negative or subcritical if it is positive:
\begin{equation}
\left. \dfrac{\partial ^3 \dot{r}}{\partial r^3}\right |_{r=0} = -3 (k_1-k_2).     
\end{equation}

For the particular case of 3 and 4-body interactions considered by Arenas and Skardal in \cite{skardal2020higher}, which corresponds to $k_2 \neq 0$, the pitchfork changes from supercritical to subcritical precisely at $k_{2} = k_1 = 2 \Delta$.

Other equilibrium points are positive real roots of the polynomial
\begin{equation} P(\rho)=(1-\rho)\left(\sum_{p=1}^{n}k_p \rho^{p-1}\right)-2\Delta,\end{equation}
which can be written explicitly as
\begin{equation} 
    P(\rho)= (k_1-2\Delta)+(k_2-k_1)\rho+\cdots+(k_n-k_{n-1})\rho^{n-1}-k_n \rho^n.
    \label{eqrho}
\end{equation}

The coefficients of a polynomial have information about its roots. According to Descartes' rule of signs, the number of positive real roots is at most the number of sign changes in the sequence of polynomial's non-zero coefficients, and the difference between the root count and the sign change count is always even. 

In the present case the coefficients are
\begin{equation} 
    (k_1-2\Delta,k_2-k_1,k_3-k_2,\cdots,k_n-k_{n-1},-k_n).
\end{equation}
Suppose the sequence of coupling constants $(k_1,...,k_n)$ is decreasing,
\begin{equation} 
    0<k_n<k_{n-1}<\cdots<k_3<k_2<k_1,
\end{equation}
which is typical of a perturbative series. 
If $k_1<2\Delta$, all coefficients are negative and we can be sure that the polynomial has no positive roots. Therefore, there are no equilibrium points other than $r=0$ and the dynamics of the system is incoherent. On the other hand, if $k_1>2\Delta$ there is one sign change in the coefficients, therefore exactly one positive real root, which branches off at $k=2\Delta$. This is the situation of the original Kuramoto model with a second-order phase transition.

Suppose now the sequence of coupling constants is increasing,
\begin{equation} k_1<k_2<\cdots<k_{n-1}<k_n.\end{equation}
If $k_1>2\Delta$, there is exactly one positive root, and if $k_1<2\Delta$ 
there are either 0 or 2 positive roots, indicating the presence of a saddle-node bifurcation and a first order phase transition in this region, as found in \cite{skardal2020higher}.

\subsection{Special cases}

\begin{figure}[H]
    \centering
    \includegraphics[width=0.6\textwidth]{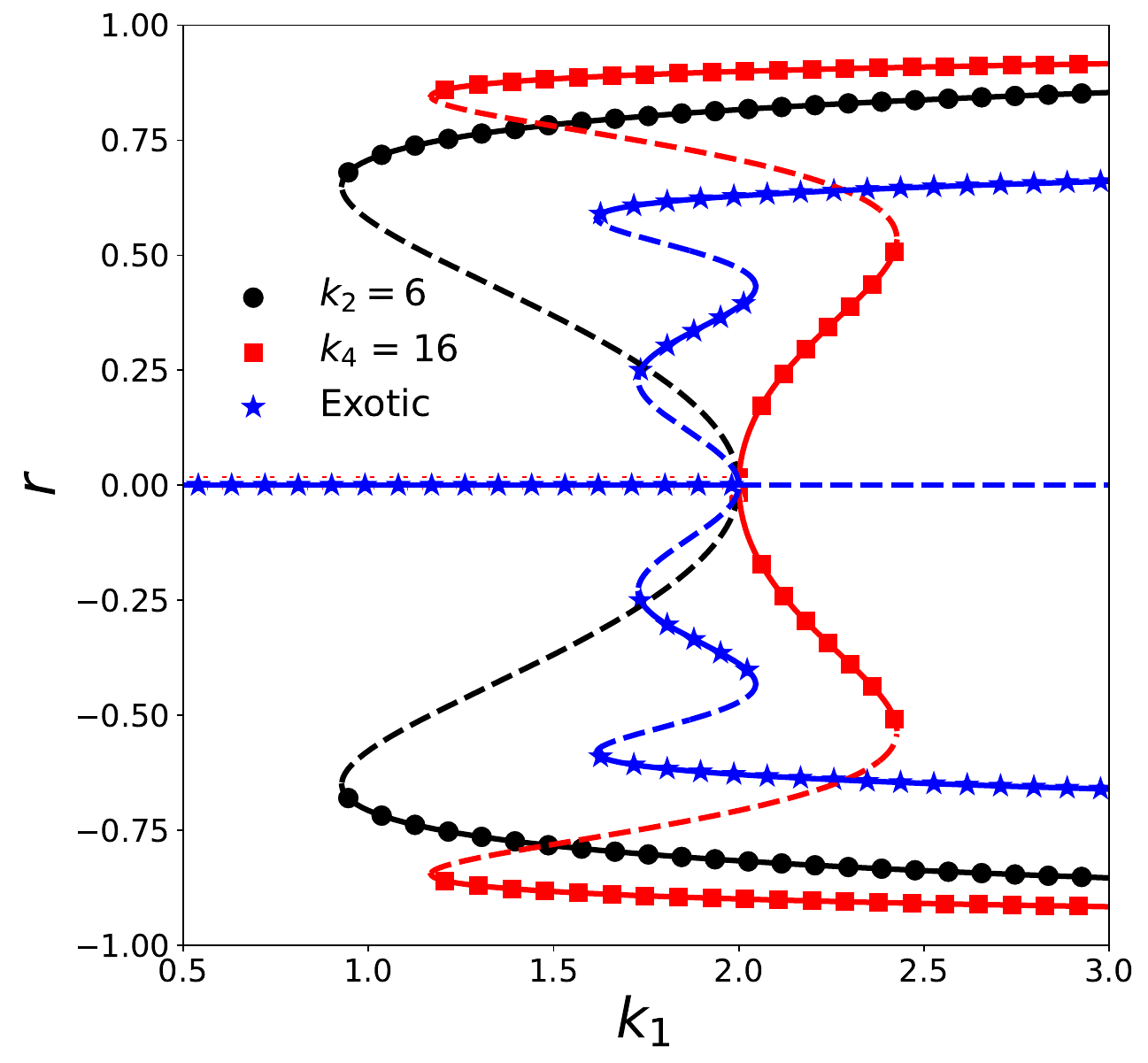}
    \caption{Example of bifurcation branches of the order parameter $r$ for some combinations of $k_n$'s. The black curve only has $k_2 = 6$; The red line corresponds to $k_4 = 16$ and the blue one is an exotic case with $k_2 = 14, k_3 = -162, k_4 = 698 $ and $k_5 = -902$. In all curves, dashed line indicates unstable branches while full lines with symbols indicate stable ones.}
    \label{fig:threestab}
\end{figure}

For arbitrary values of the coupling constants, the number of positive real roots in the interval $(0,1)$ is at most $n$, but is typically no more than two. Figure \ref{fig:threestab} shows a few examples of equilibrium points as functions of $k_1$ for $\Delta=1$. 

For $k_2=6$ and $k_n=0$ for $n>2$ (black line with circles) the pitchfork bifurcation at $k_1=2$ is subcritical and two stable solutions coexist for $0.92 \lesssim k_1 < 2.00$, in a similar fashion as in \cite{skardal2020higher}. For $k_4=16$, $k_2=k_3=0$ and $k_n=0$ for $n>4$ (red line with squares), the pitchfork bifurcation is supercritical and there is bi-stability in the interval $2.0 < k_1 \lesssim 2.4$. 



A more exotic case with $k_2=14$, $k_3=-162$, $k_4=698$, $k_5=-902$ and $k_n=0$ for $n>5$ (blue line with stars) produces a region with three stable states in the $1.75 \lesssim k_1 < 2.0$ interval. Notice, however, that the values of the coupling constants in this case are very large, quite specific and probably not realistic.

When only $k_1$ and some other $k_n$ are present, the equation determining the equilibrium solutions of the system reduce to
\begin{equation} 
(1-\rho)k_1+(1-\rho)k_n\rho^{n-1}-2\Delta = 0
\end{equation}
and we need to find the roots of
\begin{equation} 
P_n(\rho)=k_n\rho^{n}-k_n\rho^{n-1}+k_1\rho-k_1+2\Delta.
\end{equation}

The coefficients of the polynomial are $(k_n,-k_n,k_1,2\Delta-k_1)$. Assuming that $k_1$ and $k_n$ are positive, we have two sign changes if $2\Delta>k_1$ and three sign changes if $2\Delta<k_1$. Therefore, there are at most two roots if $2\Delta>k_1$ and at most three if $2\Delta<k_1$. Moreover, the difference between the number of positive real roots and the number of sign changes must be even, so if $2\Delta>k_1$, there are either 0 or 2 equilibrium states and if $2\Delta<k_1$, there are either 1 or 3 equilibrium states.

\begin{figure}[h]
    \centering
    \includegraphics[width=0.45\textwidth]{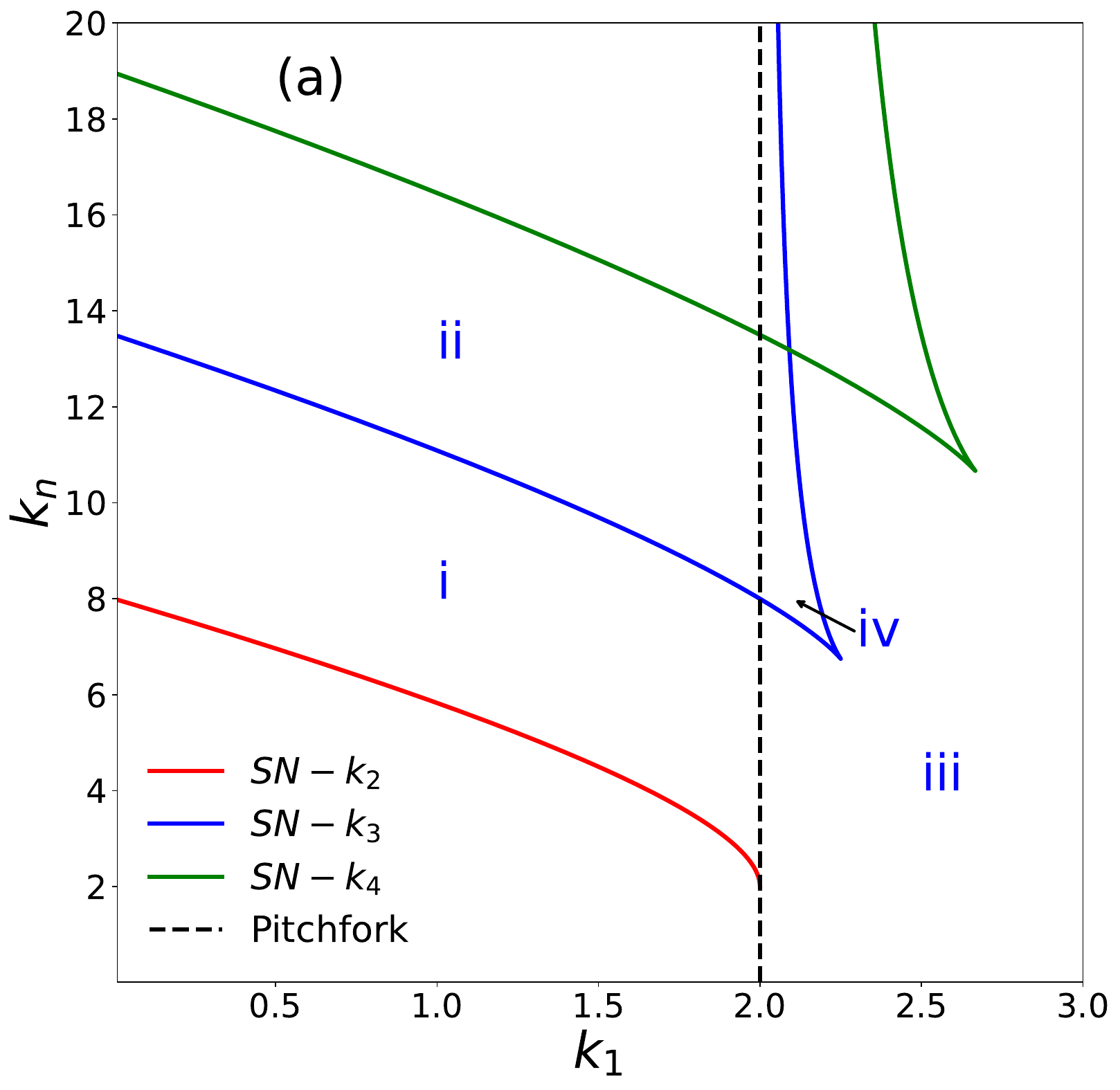}
    \includegraphics[width=0.45\textwidth]{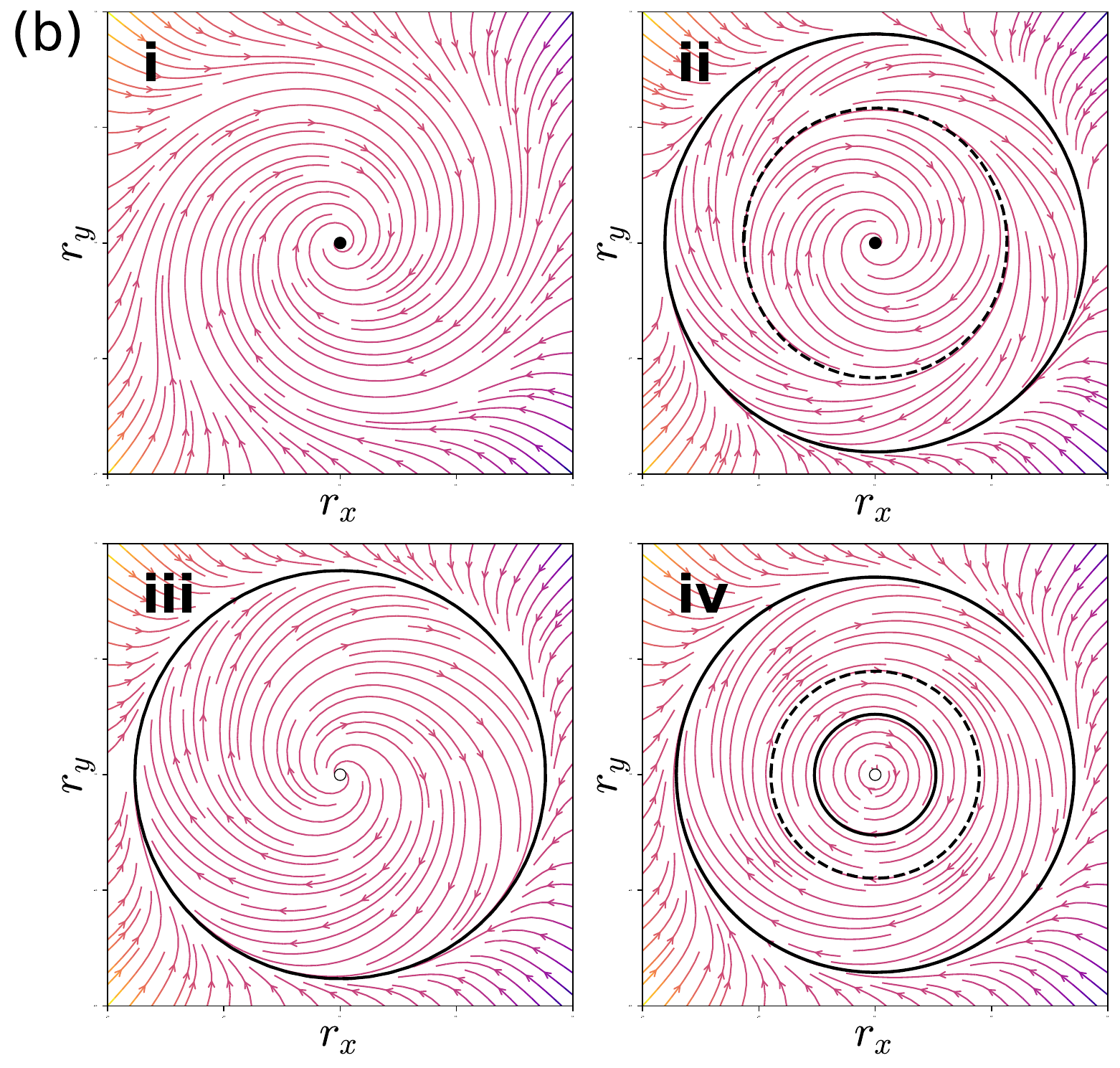}
    \caption{Bifurcation diagram in the ($k_1,k_n$) plane for different values of $n$. Roman numerals indicate the four regions into which the SN and pitchfork manifolds divide this plane for $n$ = 3. (b) Stream plots highlighting the different regions delimited by the bifurcation manifolds for $n=3$. Full symbols and full lines represent stable solutions while empty symbols and dashed lines indicate unstable ones.}
    \label{fig:bifn}
\end{figure}

We can calculate the saddle-node (SN) bifurcation manifold by solving $P_n(\rho) = 0$ and $\dfrac{\partial P_n(\rho)}{\partial \rho} = 0$ for $k_1$ and $k_n$. We find
\begin{align}
    \frac{k_1}{2\Delta} &= \frac{(n-n\rho-1)}{(n-1)(1-\rho)^2}, \\
    \frac{k_n}{2\Delta} &= \dfrac{1}{(n-1)\rho^{n-2}(1-\rho)^2}.
\end{align}
Examples of the SN manifold for $n=2$, $3$ and $4$ are shown in Fig.\ref{fig:bifn}(a). Previous works with $3$ and $4$-body interactions, such as Ref. \cite{skardal2020higher}, correspond to $n=2$, in which the SN manifold merges with the pitchfork manifold, dividing the $k_1 \times k_n$ plane into three regions: \textcolor{black}{$k_1>2$; $k_1<2$ below the red line; and $k_1<2$ above the red line, related to different asymptotic dynamics. The inclusion of even higher order interactions causes the SN manifold to stretch beyond $k_1 = 2$ and cross the pitchfork line, now dividing the $k_1 \times k_n$ plane into four regions.  Three of them are equivalent to those found in the $n = 2$ case and are labeled as regions $i$, $ii$ and $iii$ in Figure \ref{fig:bifn}(a), for the case $n = 3$.  The other corresponds to a novel behavior, labeled region $iv$, a bistable region between two synchronized states, in contrast to the synchronized-desynchronized bi-stability of region $ii$}. 


Panel 2(b) shows stream plots of the velocity field ($r_x = r\text{cos}(\psi);r_y = r\text{sin}(\psi)$) for all the four regions mentioned above, for the case $n=3$. In these panels, we set $\omega_0 = 1$ to differentiate between desyncronized (fixed points) to synchronized (periodic orbits) solutions.  Full symbols and full lines represent stable solutions while empty symbols and dashed lines indicate unstable ones.

\subsection{Correlated coupling coefficients}

Consider the case when all interactions have the same intensity, $k_1 = \cdots=k_n = k$. Then, most coefficients of Equation \eqref{eqrho} are null, reducing the characteristic polynomial to:
\begin{equation}
     P(\rho) = k -2\Delta -k\rho^n.
\end{equation}
Thus, by solving $P(\rho) = 0$ we obtain $\rho^n =  1-2\Delta/k$, or
\begin{equation}
    r^{2n} = 1-\dfrac{2\Delta}{k},
\end{equation}
which has at most one real and positive solution. 

\begin{figure}[h]
    \centering
    \includegraphics[width=0.6\textwidth]{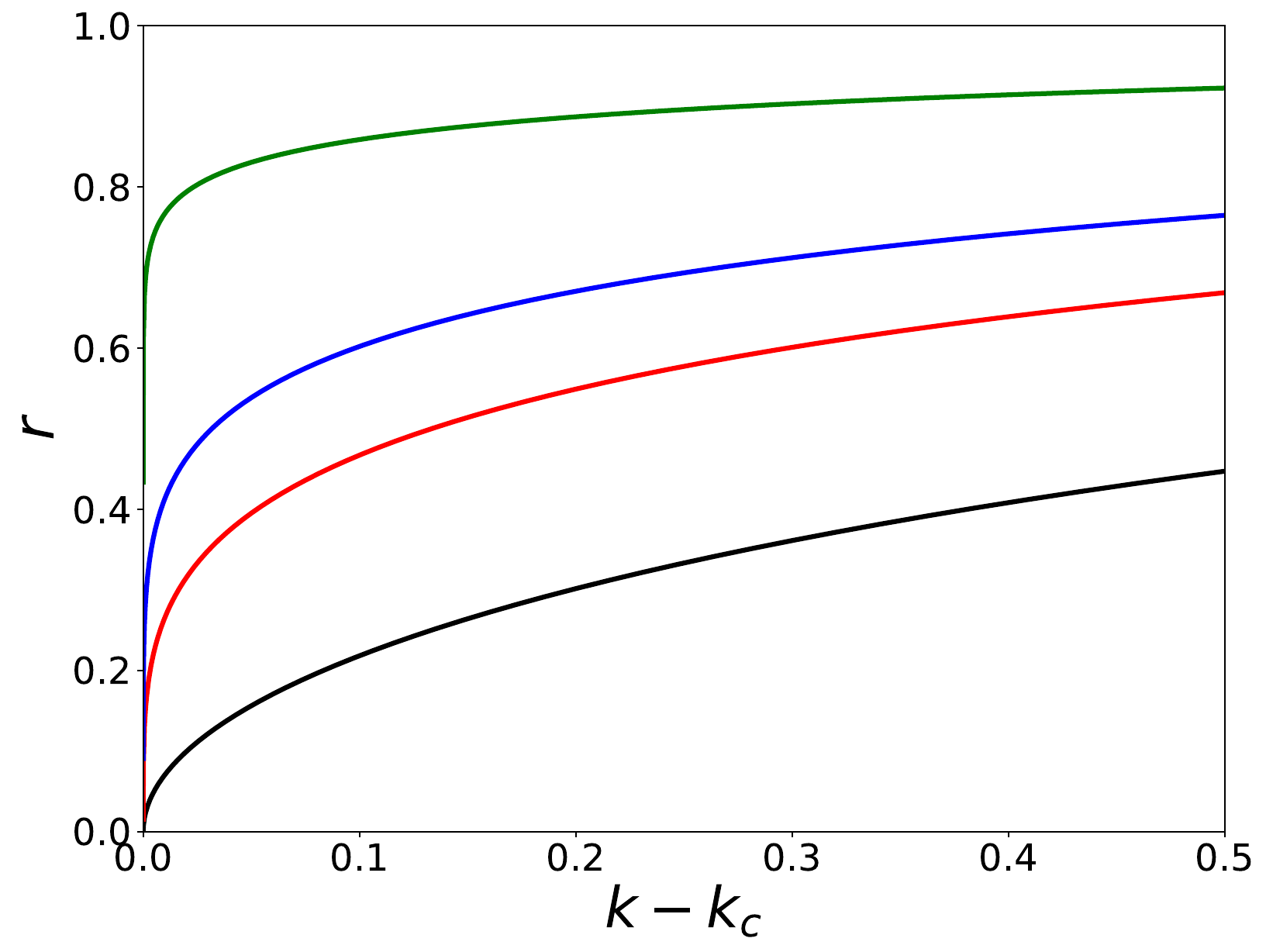}
    \caption{Equilibrium solutions for $r$ in function of $k-k_c$ ($k_c = 2\Delta$), each curve represents the highest-order $n$ with non-zero coefficient. From  bottom to top, $n=1,2,3$ and $10$. }
    \label{fig:krho}
\end{figure}

Therefore, the only bifurcation that occurs in this case is the pitchfork, at $k_c = 2\Delta$. In addition, as $n$ grows this sequence of second-order transitions approaches, as shown in Figure \ref{fig:krho}, a first-order transition, giving rise to an explosive synchronization with the order parameter abruptly going from 0 to 1 as $k$ crosses the critical value.

\begin{figure}[H]
    \centering
    \includegraphics[width=0.6\textwidth]{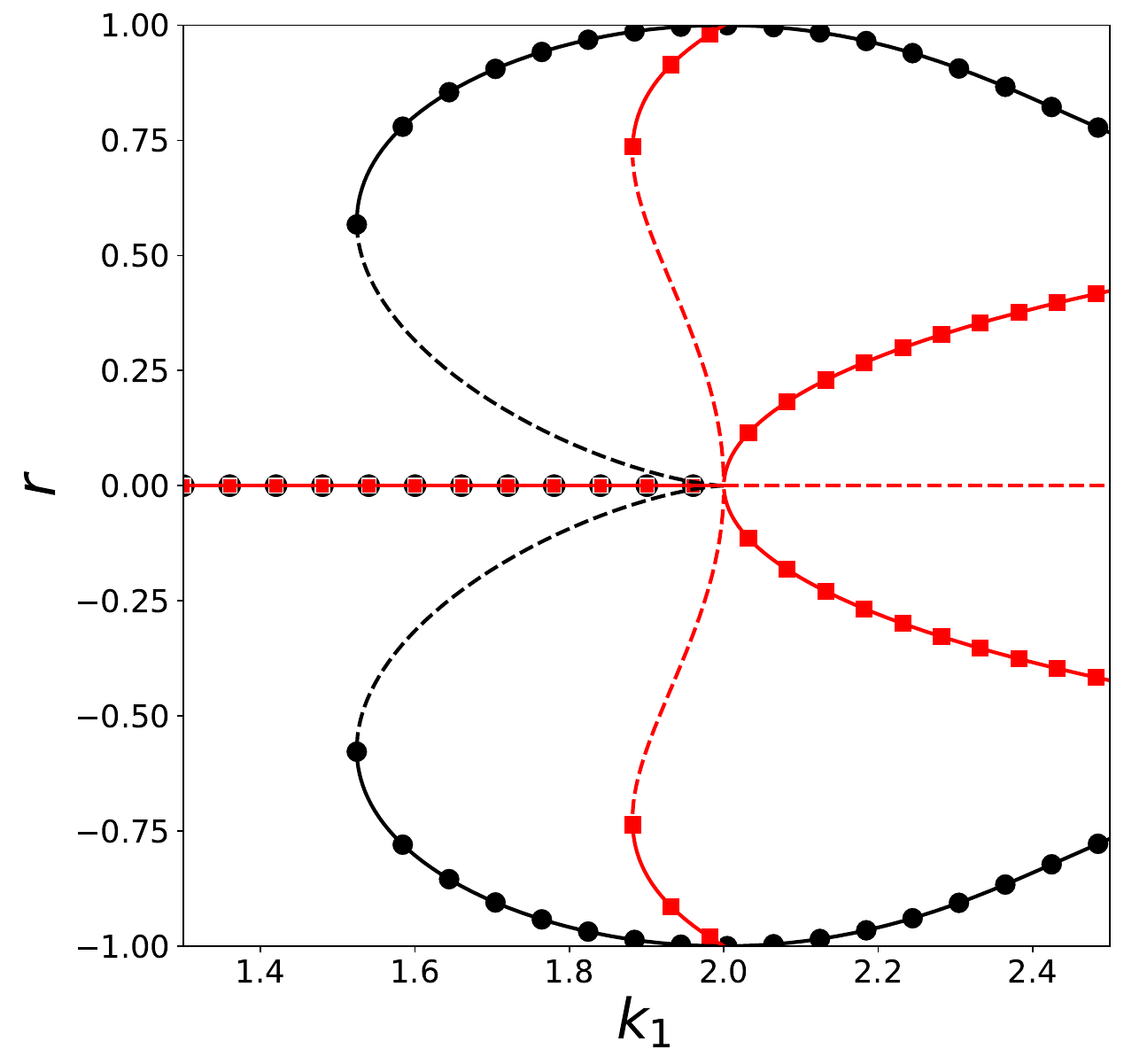}
    \caption{Equilibrium values of the order parameter as functions of $k_1$ for $k_2=(2-k_1)^{-2}$ (black line with circles) and $k_3=(2-k_1)^{-1}$ (red line with squares). In both curves, dashed line indicates unstable branches while full lines with symbols indicate stable ones.} 
    \label{fig:corre}
\end{figure}

If the relation between coupling coefficients is non-linear, the system may exhibit some very curious behavior. Let us consider, for example, the possibility that 
\begin{equation}
    k_n=\frac{1}{(2\Delta-k_1)^m}
\end{equation}
Figure \ref{fig:corre} shows two examples of equilibrium values of the order parameter as functions of $k_1$ for $\Delta=1$ and $n=m=2$ (black curve with circles) and $n=3$, $m=1$ (red curve with squares). The first case shows a classic example of hysteresis: starting close to the stable solution $r=0$ at, say, $k_1=1$ and increasing $k_1$, the solution jumps to the stable synchronized branch at $k_1=2$. Decreasing $k_1$
below 2 will keep the system in the synchronized state until $k_1 \approx 1.5$.  The bifurcation at $k_1=2\Delta$ is pitchfork subcritical. 

The second case, on the other hand, is different, as the stable synchronized branch that exists for $k_1 < 2$ will not be accessed starting ar $r=0$ and increasing or decreasing $k_1$. If the system is placed in this state, it will never go back to it once it moves away by changing $k_1$. The bifurcation is transcritical, with three branches on each side of the critical point. Writing $k_1=2+\mu$ and $k_3 = 1/(2-k_1) = -1/\mu$,  the dynamical equation close to $r=\mu=0$ can be approximated by $\dot{r} =\frac{r}{2\mu}(\mu^2 - r^4)$, which is the normal form of the bifurcation.

\section{Conclusions}
\label{conc}

Recent years have seen growing interest in interactions among physical bodies that are not pairwise but instead involve multiple bodies simultaneously \cite{sizemore2018cliques,ghosh2024chimeric,sanchez2019high,de2020social,alvarez2021evolutionary,vega2004fitness,battiston2020networks}. These are usually regarded as being of a higher order, and in network dynamics they correspond to replacing a graph model by some kind of hypergraph. The Kuramoto system, due to its simplicity, has been used as a test bed for uncovering the physical consequences of such higher order interactions \cite{skardal2020higher,dutta2023impact,leon2024higher}.

Higher order interactions can be added to the Kuramoto equations of motion in many different ways. Here we considered a very special case in which the equation for $\dot{\theta}_i$ has terms proportional to $\sin(\vec{\theta}\cdot\vec{c}-\theta_i)$, where the vector $\vec{\theta}$ represents a set of oscillators interacting with $\theta_i$ as a simplex. For this particular class of interactions it was possible to use the OA ansatz and derive reduced equations for the order parameter when taking into account all possible orders simultaneously. Our major finding was that the number of bodies involved in the interaction does not determine the contribution to the resulting dynamical equation, but rather by its effective order $p$, which depends on the $\ell_1$-norm of $\vec{c}$. In fact, the 3- and 4-body interactions studied in \cite{skardal2020higher} had the same $p$ and led to the same dynamical equations.

We also derived some general results about the bifurcations that occur in these systems in terms of the coupling constants and considered some special cases of interest. In particular, we have found that interactions with more than 4 bodies may exhibit bi-stability between two synchronized states. Finally, we have found some curious scenarios for specific parameter choices, such as a tri-stability case and non usual discontinuous transitions. We leave the exploration of other functional higher order interactions for future work.
\begin{acknowledgments}
	 This work was partly supported by FAPESP, grants 2021/14335-0 (MAMA), 2023/03917-4 (GSC) and 2023/15644-2 (MN), and also by CNPq, grants 303814/2023-3 (MAMA) and 304986/2022-4 (MN).
\end{acknowledgments}


\end{document}